\documentstyle[12pt,a4]{article}

\begin{document}

\textheight 21.7 true cm
\textwidth 14.1 true cm

\voffset -0.7 true cm
\hoffset 0.5 true cm

\baselineskip 16pt
\parskip 8pt
\parindent 28pt

\title{Does a non-zero tunnelling probability imply particle 
production in time independent classical electromagnetic 
backgrounds?} 
\author{L.~Sriramkumar\thanks{lsk@iucaa.ernet.in}\quad and \quad 
T.~Padmanabhan\thanks{paddy@iucaa.ernet.in}\\
IUCAA, Post Bag 4, Ganeshkhind, Pune 411 007, INDIA.} 
\maketitle

\begin{abstract}

In this paper, we probe the validity of the tunnelling interpretation 
that is usually called forth in literature to explain the phenomenon
of particle production by time independent classical electromagnetic 
backgrounds.  
We show that the imaginary part of the effective lagrangian is zero 
for a complex scalar field quantized in a time independent, but 
otherwise arbitrary, magnetic field. 
This result implies that no pair creation takes place in such 
a background. 
But we find that when the quantum field is decomposed into its normal 
modes in the presence of a spatially confined and time independent magnetic 
field, there exists a non-zero tunnelling probability for the effective 
Schr{\" o}dinger equation.
According to the tunnelling interpretation, this result would imply that 
spatially confined magnetic fields can produce particles, thereby 
contradicting the result obtained from the effective lagrangian.
This lack of consistency between these two approaches calls into 
question the validity of attributing a non-zero tunnelling probability 
for the effective Schr{\" o}dinger equation to the production of particles 
by the time independent electromagnetic backgrounds. 
The implications of our analysis are discussed.
\end{abstract} 
\newpage

\section{Introduction}\label{sec:intro}

The phenomenon of pair creation by classical electromagnetic 
backgrounds was first studied by Schwinger more than four 
decades ago.
In his classic paper, Schwinger considered a quantized spinor 
field interacting with a constant external electromagnetic 
background~\cite{schwinger51}.
Obtaining an effective lagrangian by integrating out the degrees of 
freedom corresponding to the quantum field, he showed that the effective 
lagrangian had an imaginary part only when $({\bf E}^2-{\bf B}^2)>0$, 
where~${\bf E}$ and~${\bf B}$ are the constant electric and magnetic 
fields respectively (also see~\cite{hande36}).
The appearance of an imaginary part in the effective lagrangian 
implies an instability of the vacuum and Schwinger attributed the 
cause of this instability to the production of pairs corresponding 
to the quantum field by the electromagnetic background.
The imaginary part of the effective lagrangian, Schwinger concluded, 
should be interpreted as the number of pairs that have been produced, 
per unit four-volume, by the external electromagnetic field. 

Though attempts have been made in literature to obtain the effective 
lagrangian for a fairly non-trivial electromagnetic field~(see for 
instance refs.~\cite{stephens89,brezin70,mandv88,mandv89,fry95}), 
its evaluation for an arbitrary vector potential proves to be 
an uphill task.
Due to this reason the phenomenon of particle production 
in classical electromagnetic backgrounds has been repeatedly 
studied in literature by the method of normal mode 
analysis.
In this approach, the normal modes of the quantum field are obtained 
by solving the wave equation it satisfies for the given electromagnetic 
background in a particular gauge. 
The coefficients of the positive frequency normal modes of the quantum
field are then identified to be the annihilation operators.
The evolution of these operators therefore follow the evolution of 
the normal modes. 
Then, by relating these operators defined in the asymptotic regions 
(either in space or in time) the number of particles that have been 
produced by the electromagnetic background can be computed.

Consider an electromagnetic background that can be represented by a 
time {\it dependent} gauge. 
If we choose to study the evolution of the quantum field in such a gauge, 
then a positive frequency normal mode of the quantum field at late times
will, in general, prove to be a linear superposition of the positive and 
negative frequency modes defined at early times.
The coefficients in such a superposition are the Bogolubov coefficients 
$\alpha$ and $\beta$. 
A non-zero Bogolubov coefficient~$\beta$ would then imply that the 
{\it in}-vacuum state is not the same as the {\it out}-vacuum state.
This in turn implies that the {\it in}-{\it out} transition 
amplitude is less than unity which can be attributed to the 
excitation of the modes of the quantum field by the electromagnetic 
background~\cite{narozhnyi68,nikishov70one,nandn70,grib72,popov72,nandn76}.
These excitations manifest themselves as real particles corresponding 
to the quantum field.

On the other hand, consider an electromagnetic background that can be 
described by a space dependent gauge (by which we mean a gauge that is 
completely independent of time).
If the evolution of the quantum field is studied in such a gauge, then 
due to the lack of dependence on time, the Bogolubov coefficient $\beta$ 
proves to be trivially zero.
This could then imply that the electromagnetic background which is 
being considered does not produce particles.

An interesting situation arises when the same electromagnetic field can be
described by a (purely) space dependent gauge as well as a (purely) time
dependent gauge.
If we choose to study the evolution of the quantum field in the time 
dependent gauge, in general, $\beta$ will prove to be nonzero thereby 
implying (as discussed above) that particles are being produced by the 
electromagnetic background. 
But, in the space dependent gauge $\beta$ is trivially zero thereby
disagreeing with result obtained in the time dependent gauge.
Therefore, to obtain results that are gauge invariant, the phenomenon of 
particle production has to be somehow `explained' in the space dependent 
gauge.  
In literature, a `tunnelling interpretation' is usually invoked to explain 
the phenomenon of particle production in such a
situation~\cite{nikishov70two,itzykson80,wang88,fulling89,brout95two}.
In this approach, an effective Schr{\" o}dinger equation is obtained 
after the quantum field is decomposed into normal modes in the space 
dependent gauge.
The non-zero tunnelling probability for this Schr{\" o}dinger equation
is then attributed  to the production of particles by the electromagnetic 
background.

The discussion in the above paragraph can be illustrated by the
following well known, but instructive, example.
Consider a constant electric field given by ${\bf E}=E\,{\hat {\bf x}}$, 
where $E$ is a constant and ${\hat {\bf x}}$ is the unit vector along the
positive $x$-axis.
This electric field can be described either by the time dependent 
gauge $A_1^{\mu}=(0, -Et, 0, 0)$ or by the space dependent gauge 
$A_{2}^{\mu}=(-Ex, 0, 0, 0)$.
In the gauge $A_1^{\mu}$, due to the time dependence, the positive 
frequency normal modes of the quantum field at $t=+\infty$ are related 
by a non-zero Bogolubov coefficient $\beta$ to the positive frequency 
modes at $t=-\infty$.
The quantity ${\vert \beta\vert}^2$ then yields the number of particles 
that have been produced in a single mode of the quantum field at late 
times in the {\it in}-vacuum~\cite{paddy91,brout91}.
But, if the evolution of the quantum field is studied in the gauge
$A_2^{\mu}$, because of time independence, $\beta$ proves to be zero 
thereby disagreeing with the result obtained in the gauge $A_1^{\mu}$.
The tunnelling interpretation can be invoked in such a situation to
explain particle production in the gauge $A_2^{\mu}$.
In this gauge, after the normal mode decomposition of the quantum 
field, an effective Schr{\" o}dinger equation is obtained along the
$x$-direction.
The non-zero tunnelling probability, ${\vert T\vert}^2$, for this 
Schr{\" o}dinger equation is then interpreted as the number of particles 
that have been produced in a single mode of the quantum 
field~\cite{paddy91,brout91}.
The tunnelling probability ${\vert T\vert}^2$ evaluated in the gauge
$A_{2}^{\mu}$, in fact, exactly matches the quantity ${\vert \beta \vert}^2$ 
obtained in the gauge $A_{1}^{\mu}$. 
Also, these two quantities agree with the pair creation rate obtained 
by Schwinger from the imaginary part of the effective lagrangian.

The fact that the quantities ${\vert \beta \vert}^2$ and 
${\vert T \vert}^2$ agree, not only with each other but 
also with the pair creation rate obtained from the effective 
lagrangian, for the case of a constant electric field has 
given certain credibility to the tunnelling interpretation.
Our aim, in this paper, is to probe the validity of the tunnelling 
interpretation. 

Consider an arbitrary electromagnetic background that can be described
by a space dependent gauge.
Also assume that when the evolution of the quantum field is studied in 
such a gauge, there exists a non-zero tunnelling probability for the
effective Schr{\" o}dinger equation.
Can such a non-zero tunnelling probability be always interpreted as 
particle production?
We attempt to answer this question in this paper by comparing 
the results obtained from the effective lagrangian with those 
obtained from the tunnelling approach.
We carry out our analysis for a spatially varying, time independent 
magnetic field when it is described by a space dependent gauge.
We find that there exists---in general---a lack of consistency between 
the results obtained  from  the tunnelling approach and those obtained 
from the effective lagrangian.
This inconsistency clearly calls into question the validity of the 
tunnelling interpretation as it is presently understood in literature.

This paper is organized as follows. 
In section~\ref{sec:efflag}, we show that the imaginary part of the 
effective lagrangian for an arbitrary time independent background 
magnetic field is zero. 
In section~\ref{sec:tunnprob}, we calculate the tunnelling probability,
which happens to be non-zero, for a particular spatially confined and 
time independent magnetic field when it is represented by a space
dependent gauge.
Finally, in section~\ref{sec:concl} we discuss the implications of our 
analysis to the study of particle production in time independent 
electromagnetic and gravitational backgrounds.

\section{Effective lagrangian for a time independent magnetic field
background}\label{sec:efflag}

The system we consider in this paper consists of a complex scalar 
field $\Phi$ interacting with an electromagnetic field represented 
by the vector potential $A^{\mu}$. 
It is described by the lagrangian density
\begin{equation} 
{\cal L}(\Phi, A_{\mu})
=\left(\partial_{\mu}\Phi+iqA_{\mu}\Phi\right) 
\left(\partial^{\mu} \Phi^* - iq A^{\mu}\Phi^*\right)
-m^2\Phi\Phi^*
- {1 \over 4} F^{\mu\nu}F_{\mu\nu},\label{eqn:l}
\end{equation}
where $q$ and $m$ are the charge and the mass associated with a 
single quantum of the complex scalar field, the asterisk denotes 
complex conjugation and 
\begin{equation}
F_{\mu\nu} = \partial_{\mu}A_{\nu} - \partial_{\nu} A_{\mu}.
\end{equation}
The electromagnetic field is assumed to behave classically, hence
$A_{\mu}$ is just a $c$-number while the complex scalar  field is 
assumed to be a quantum field so that $\Phi$ is an operator valued 
distribution.
We will also assume that the electromagnetic field is given to us
apriori, {\it i.e.} we will not take into account the backreaction of 
the quantum field on the classical background.
(Kiefer {\it et. al.} show in ref.~\cite{kiefer91} that the 
semiclassical domain as envisaged here does exist; 
also see ref.~\cite{kiefer92} in this context.
The issue of backreaction on the electromagnetic background has been 
addressed in refs.~\cite{cooper89,kluger92,brout95one}.) 
In such a situation, we can obtain an effective lagrangian for the 
classical electromagnetic background by integrating out the degrees 
of freedom corresponding to the quantum field as follows
\begin{equation}
\exp \,i\int d^4x\,{\cal L}_{eff}(A_{\mu})
\equiv\int{\cal D}\Phi\, \int{\cal D}\Phi^*\,
\exp \,i\int d^4x\,
{\cal L}(\Phi, A_{\mu}),\label{eqn:leff}
\end{equation}
where we have set $\hbar=c=1$ for convenience.
The effective lagrangian can be expressed as
\begin{equation}
{\cal L}_{eff}= {\cal L}_{em}+{\cal L}_{corr},
\end{equation}
where ${\cal L}_{em}$ is the lagrangian density for the free 
electromagnetic field, the third term in the lagrangian
density~(\ref{eqn:l}), and ${\cal L}_{corr}$ is given by
\begin{eqnarray}
& &\exp \,i\int d^4x\,{\cal L}_{corr}(A_{\mu})\nonumber\\
& &\qquad\quad\quad
= \int{\cal D}\Phi\, \int {\cal D}\Phi^*\exp \,i\int d^4x\,
\biggl\lbrace\left(\partial_{\mu}\Phi+iqA_{\mu}\Phi\right) 
\left(\partial^{\mu} \Phi^* - iq A^{\mu}\Phi^*\right)\nonumber\\
& &\qquad\qquad\qquad\qquad\qquad\qquad\qquad\qquad\qquad\qquad
\qquad\qquad\quad\quad
-m^2\Phi\Phi^*\biggl\rbrace.
\end{eqnarray}
Integrating the action for the scalar field in the above equation
by parts and dropping the resulting surface terms, we obtain that 
\begin{equation}
\exp\,i\int d^4x\,{\cal L}_{corr}(A_{\mu})= 
\int{\cal D}\Phi\,\int{\cal D}\Phi^*\, 
\exp -i\int d^4x\;\Phi^*{\hat D}\Phi
=\left({\rm det}\,{\hat D}\right)^{-1},\label{eqn:lcorr1}
\end{equation}
where the operator ${\hat D}$ is given by 
\begin{equation}
{\hat D}\equiv D_{\mu}D^{\mu}+m^2 \qquad {\rm and}
\qquad D_{\mu}\equiv\partial_{\mu}+iqA_{\mu}.\label{eqn:D}
\end{equation}

The determinant in equation~(\ref{eqn:lcorr1}) can be expressed as 
follows
\begin{equation}
\exp i \int d^4x\, {\cal L}_{corr}
= \left({\rm det}\,{\hat D}\right)^{-1}
=\exp -\,{\rm Tr}(\ln{\hat D})
=\exp -\int d^4x\, 
\langle t, {\bf x}\vert \,\ln {\hat D}\,\vert t, {\bf x}\rangle,
\end{equation}
and in arriving at the last expression, following Schwinger, we have 
chosen the set of basis vectors $\vert t, {\bf x}\rangle$ to evaluate 
the trace of the operator $\ln {\hat D}$. 
From the above equation it is easy to identify that
\begin{equation}
{\cal L}_{corr}= i\, \langle t, {\bf x}\vert\ln {\hat D}
\vert t, {\bf x}\rangle.
\end{equation}
Using the following integral representation for the operator 
$\ln {\hat D}$, 
\begin{equation}
\ln {\hat D}\equiv -\int_0^{\infty}{ds \over s}\, 
\exp -i({\hat D}-i\epsilon)s\label{eqn:lnD}
\end{equation}
(where $\epsilon \rightarrow 0^{+}$), the expression for 
${\cal L}_{corr}$ can be written as 
\begin{equation}
{\cal L}_{corr}=-i\,\int_{0}^{\infty} {ds \over s}\,
e^{-i(m^2-i\epsilon)s}\,
K(t, {\bf x}, s\vert t, {\bf x}, 0),\label{eqn:lcorr2}
\end{equation}
where 
\begin{equation}
K(t, {\bf x}, s\,\vert\, t, {\bf x}, 0)
=\langle t, {\bf x} \vert\, e^{-i{\hat H}s}\,\vert t, {\bf x}\rangle
\qquad {\rm and}\qquad
{\hat H}\equiv D_{\mu}D^{\mu}.
\end{equation}
That is, $K(t, {\bf x}, s\,\vert\, t, {\bf x}, 0)$ is the kernel 
for a quantum mechanical particle (in 4~dimensions) described 
by the hamiltonian operator ${\hat H}$.
The variable $s$, that was introduced in~(\ref{eqn:lnD}) when 
the operator $\ln{\hat D}$ was expressed in an integral form,
acts as the time parameter for the quantum mechanical system.
(The integral representation for the operator $\ln{\hat D}$ we 
have used above is divergent in the lower limit of the integral, 
{\it i.e.} near $s=0$.
This divergence is usually regularized in field theory by 
subtracting from it another divergent integral, {\it viz.} 
the integral representation of an operator $\ln{\hat D_0}$, 
where ${\hat D_0} =(\partial^{\mu}\partial_{\mu}+m^2)$, the 
operator corresponding to that of a free quantum field. 
That is, to avoid the divergence, the integral representation 
for $\ln {\hat D}$ is actually considered to be
\begin{equation}
\ln {\hat D} -\ln {\hat D_0}\equiv 
-\int_0^{\infty}{ds \over s}\, 
\left(\exp-i({\hat D}-i\epsilon)s\;
-\;\exp-i({\hat D_0}-i\epsilon)s\right).
\end{equation}
Therefore, in what follows, the operator $\ln{\hat D}$ should be 
considered as $\ln{\hat D}-\ln{\hat D_0}$ though it will not be 
written so explicitly.)

Now, consider a background electromagnetic field described by the 
vector potential
\begin{equation}
A^{\mu}=(0, 0, A(x), 0),\label{eqn:A1}
\end{equation}
where $A(x)$ is an arbitrary function of $x$.
This vector potential does not produce an electric field but gives 
rise to a magnetic field ${\bf B}= (dA/dx)\, {\hat {\bf z}}$, where 
${\hat{\bf z}}$ is the unit vector along the positive $z$-axis. 
According to the Maxwell's equations, in the absence of an electric
field, the magnetic field is related to the current ${\bf j}(x)$ as
follows 
\begin{equation}
{\bf \nabla}\times {\bf B} ={\bf j}.\label{eqn:max}
\end{equation}
Then, the current that can give rise to the time independent magnetic 
field we consider here is given by
\begin{equation}
{\bf j}=-\left({d^2 A\over dx^2}\right)\,{\hat {\bf y}},
\end{equation}
where ${\hat{\bf y}}$ is the unit vector along the positive $y$-axis.
If we assume that ${\bf j}$ is finite and continuous everywhere and also 
vanishes as $\vert x \vert\rightarrow\infty$, then the magnetic field we 
consider here can be physically realised in the laboratory.

The operator ${\hat H}$ corresponding to the vector 
potential~(\ref{eqn:A1}) is given by
\begin{equation}
{\hat H}\equiv {\partial_{t}}^2 - {\nabla}^2 + 2iq A \partial_y +q^2 A^{2}.
\end{equation}
Then, the kernel for the quantum mechanical particle described by  
the hamiltonian above can be formally written as 
\begin{equation}
K(t, {\bf x}, s\,\vert\, t,{\bf x}, 0) 
= \langle t, {\bf x} \vert \exp -i({\partial_{t}}^2 - {\nabla}^2 
+ 2iq A \partial_{y} +q^2 A^2) s\,\vert t, {\bf x}\rangle.
\end{equation}
Using the translational invariance of the hamiltonian operator 
${\hat H}$ along the time coordinate $t$ and the spatial coordinates 
$y$ and $z$, we can express the above kernel as follows 
\begin{equation} 
K(t, {\bf x}, s\,\vert\, t, {\bf x}, 0) 
=\int {d\omega \over 2\pi} \int {dp_{y} \over 2\pi}
\int {dp_{z} \over 2\pi}\; \langle x\vert\, 
\exp -i(-\omega^2-d_{x}^2+(p_{y}-qA)^2+{p_{z}}^2)s\, \vert x \rangle.
\end{equation}
Performing the $\omega$ and $p_z$ integrations, we obtain that
\begin{equation}
K(t, {\bf x}, s\,\vert\, t, {\bf x}, 0)
={1 \over 4\pi s}\int {dp_y \over 2\pi}\;
\langle x\vert \, e^{-i{\hat G}s}\, \vert x \rangle
\qquad{\rm where}\qquad
{\hat G}\equiv -d_x^2+(p_y-qA)^2.
\end{equation}
The quantity $\langle x\vert \, e^{-i{\hat G}s}\, \vert x \rangle$ 
is then the kernel for the one dimensional quantum mechanical system 
described by the effective hamiltonian operator ${\hat G}$.
It can expressed, using the Feynman-Kac formula, as
\begin{equation}
\langle x\vert \, \exp{-i{\hat G}s}\, \vert x \rangle
= \sum_{E} {\vert \Psi_E(x)\vert}^2\, e^{-iEs},
\end{equation}
where $\Psi_E$ is the eigenfunction of the operator ${\hat G}$
corresponding to an eigenvalue $E$, {\it i.e.}
\begin{equation}
{\hat G}\Psi_E\equiv (-d_x^2+(p_y-qA)^2)\Psi_E=E\Psi_{E},
\end{equation}  
so that $K(t, {\bf x}, s\,\vert\, t, {\bf x}, 0)$ reduces to
\begin{equation}
K(t, {\bf x}, s\,\vert\, t, {\bf x}, 0)
={1 \over 4\pi s}\int {d p_y \over 2 \pi}
\sum_{E} {\vert \Psi_E(x)\vert}^2\, e^{-iEs}.
\end{equation}
(It is assumed that the summation over $E$ stands for integration over the 
relevant range when $E$ varies continuously.)
Since the potential term, $(p_y-qA(x))^2$, in the hamiltonian operator 
${\hat G}$ is a positive definite quantity, the eigenvalue $E$ can only 
lie in the range $(0, \infty)$.
Substituting the expression for $K(t, {\bf x}, s \,\vert\, t, {\bf x}, 0)$
in~(\ref{eqn:lcorr2}), we find that ${\cal L}_{corr}$ is given by
\begin{equation}
{\cal L}_{corr}= -{i \over 4\pi}\int{dp_y \over 2\pi}\,
\sum_{E} {\vert \Psi_E(x)\vert}^2\, \int_0^{\infty} {ds \over s^2}\,
{e^{-i(m^2+E-i\epsilon)s}}.
\end{equation}
Differentiating the above expression for ${\cal L}_{corr}$ twice 
with respect to $m^2$ (cf.~\cite{landau4}) and then carrying out 
the integration over the variable $s$, we obtain that
\begin{equation}  
{\cal L}_{corr}^{''}
={\partial^2 {\cal L}_{corr} \over {\partial (m^2)^2}}
= {1 \over 4\pi}\int{dp_y \over 2\pi}\,
\sum_{E} \left({{\vert \Psi_E(x) \vert}^2 \over 
{m^2+E-i\epsilon}}\right).\label{eqn:ddlcorr}
\end{equation}
The quantity $(m^2+E-i\epsilon)^{-1}$ in the above expression, can be 
written as
\begin{equation}
\left({1 \over m^2+E-i\epsilon}\right)={\cal P}\left({1 \over m^2+E}\right)
+i\pi\, \delta(m^2+E),
\end{equation}
where ${\cal P}$ is the principal value of the corresponding argument.
Since $E$ is a positive semi definite quantity, the argument of the 
$\delta$-function above never reduces to zero.
Therefore the  second term in the above expression vanishes with the 
result that ${\cal L}_{corr}^{''}$ is a real quantity thereby implying 
that ${\cal L}$ is also a real quantity. 
In fact, integrating ${\cal L}_{corr}^{''}$ twice with respect to $m^2$, 
we find that ${\cal L}_{corr}$ can be expressed as
\begin{equation}
{\cal L}_{corr}= {1 \over 4\pi}\int{dp_y \over 2\pi}\,
\sum_{E} {\vert \Psi_E(x)\vert}^2\, \alpha\, (\ln \alpha -1),
\end{equation}
where $\alpha =(m^2+E)>0$ and $\epsilon$ has been set to zero. 
Then, clearly ${\cal L}_{corr}$ is a real quantity.
(To be rigorous, one has to take into account the two constants of 
integration that will appear on integrating ${\cal L}_{corr}^{''}$ with 
respect to $m^2$ (see~\cite{landau4}), but these constants are irrelevant 
for our arguments here.)

Though we are unable to evaluate the effective lagrangian for an 
arbitrary time independent magnetic field in a closed form, we have 
been able to show that it certainly does not have an imaginary part.
Therefore we can unambiguously conclude that time independent 
background magnetic fields do not produce particles.
This, of course agrees with Schwinger's result for a constant (time 
independent) magnetic field background. 

\section{Tunneling probability in a time independent magnetic field
background}\label{sec:tunnprob}

We shall now calculate the tunnelling probability for the a specific time 
independent background magnetic field in a space dependent gauge.
Consider the vector potential 
\begin{equation}
A^{\mu}=(0, 0, B_0\, L\, \tanh(x/L), 0),\label{eqn:A2}
\end{equation}
where $B_0$ and $L$ are arbitrary constants.
This vector potential does not produce an electric field but gives rise 
to the following magnetic field  
\begin{equation}
{\bf B}=B_0\, {\rm sech}^2(x/L)\; {\hat{\bf z}},\label{eqn:mf}
\end{equation}
where ${\hat{\bf z}}$ is the unit vector along the positive $z$-axis.
The magnetic field ${\bf B}$ goes to zero as $\vert x\vert\rightarrow 
\infty$, {\it i.e} its strength is confined to an effective width $L$ 
along the $x$-axis.
In the absence of an electric field, according to the Maxwell's 
equation~(\ref{eqn:max}), the magnetic field given by~({\ref{eqn:mf}) 
can be produced by the current 
\begin{equation}
{\bf j}=\left({2 B_0\over L}\right)\, 
{\rm sech}(x/L)\, \tanh(x/L)\, {\hat {\bf y}},  
\end{equation}
where, as before, ${\hat{\bf y}}$ denotes the unit vector along the 
positive $y$-axis.
The current ${\bf j}$ is finite and continuous everywhere and also goes 
to zero as $\vert x\vert \rightarrow \infty$. 
Therefore the magnetic field ${\bf B}$ given by~(\ref{eqn:mf}) is 
physically realisable in the laboratory.

In an electromagnetic background, described by the vector potential 
$A^{\mu}$, the complex scalar field satisfies the following Klein-Gordon 
equation 
\begin{equation}
(D_{\mu}D^{\mu} +m^2)\Phi= 
(\partial_{\mu}+iqA_{\mu})\,(\partial_{\mu}+iqA_{\mu})\Phi=0.
\end{equation}
Substituting the vector potential~(\ref{eqn:A2}) in the 
above equation, we obtain that
\begin{equation}
(\partial_t^2-\nabla^2 + 2iqB_0 L \tanh(x/L) \partial_y
+q^2B_0^2L^2 \tanh^2(x/L) +m^2)\Phi=0.\label{eqn:kg}
\end{equation}
Since the vector potential~(\ref{eqn:A2}) is dependent only on the 
spatial coordinate $x$, the normal mode decomposition of the scalar 
field can be carried out as follows 
\begin{equation} 
\Phi_{\omega k_{\perp}}=N_{\omega k_{\perp}}
e^{-i\omega t}\, e^{i {\bf k}_{\perp}.{\bf x}_{\perp}}\, 
\psi_{\omega k_{\perp}}(x),\label{eqn:nm}
\end{equation}
where $N_{\omega k_{\perp}}$ is the normalisation constant, 
$k_{\perp}\equiv(k_y, k_z)$ and $x_{\perp}\equiv(y, z)$ .
The modes are normalized according to the scalar product 
\begin{eqnarray}
(\Phi_{\omega k_{\perp}}, \Phi_{\omega' {k'}_{\perp}})
&=& -i \int d\Sigma^{\mu} \left(\Phi_{\omega k_{\perp}}
(D_{\mu} \Phi_{\omega' {k'}_{\perp}})^*
-\Phi_{\omega' {k'}_{\perp}}^*
(D_{\mu}\Phi_{\omega k_{\perp}})\right)\nonumber\\ 
&=& \delta(\omega-\omega')\,
\delta(k_{\perp}-{k'}_{\perp}),
\end{eqnarray}
where $d\Sigma^{\mu}$ is a timelike hypersurface.
Substituting the normal mode $\Phi_{\omega k_{\perp}}$  
in~(\ref{eqn:kg}), we find that $\psi$ satisfies the following 
differential equation 
\begin{equation}
{d^2\psi \over d\rho^2}+\left(\omega^2-(k_y - q B_0 L \tanh{\rho})^2
-k_z^2-m^2\right)\,L^2 \,\psi = 0\label{eqn:depsi} 
\end{equation}
where $\rho=(x/L)$ and we have dropped the subscripts on $\psi$.
This differential equation can be rewritten as
\begin{equation}
-{d^2\psi\over d\rho^2}+(k_y L-q B_0 L^2 \tanh\rho)^2\, \psi
=(\omega^2-k_z^2-m^2)\,L^2\,\psi,\label{eqn:effsch}
\end{equation}
which then resembles a time independent Schr{\" o}dinger equation
corresponding to a potential ${(k_y L -q B_0 L^2 \tanh\rho)^2}/2$ 
and energy eigenvalue $(\omega^2-k_z^2-m^2)L^2/2$.
The potential term in the effective Schr{\" o}dinger equation above
reduces to a finite constant as $\vert x \vert \rightarrow \infty$.
Therefore, there exist solutions for $\psi$ which reduce 
to $e^{\pm i k_L x}$ as $x\rightarrow -\infty$ and $e^{\pm i k_R x}$ 
as $x\rightarrow +\infty$, where $k_L$ and $k_R$ are given by 
\begin{eqnarray}
k_L &=& \left(\omega^2-(k_y+q B_0 L)^2
-k_z^2-m^2\right)^{1/2},\nonumber\\
k_R &=& \left(\omega^2-(k_y-q B_0 L)^2
-k_z^2-m^2\right)^{1/2}.\label{eqn:klkr}
\end{eqnarray}
We will confine to values of $\omega$ and $k_{\perp}$ such that $k_L$ 
and $k_R$ are real.

The differential equation~(\ref{eqn:depsi}) can be solved by the 
following ansatz~(cf.~\cite{mandf53})
\begin{equation}
\psi= e^{-a\rho}\, {\rm sech}^{b}\rho\, 
f(\rho)\label{eqn:psi}
\end{equation}
where 
\begin{equation}
a=ik_{-}L\quad;\quad b=ik_{+}L\quad {\rm and}\quad k_{\pm}= (k_R \pm k_L)/2.
\end{equation}
Substituting the above ansatz in~(\ref{eqn:depsi}), we find 
that $f$ satisfies the following differential equation 
\begin{equation}
u(u-1)\,{d^2f \over du^2} + (1+a+b - 2(b+1)u)\, {df \over du} 
+ (q^2 {B_0}^2 L^4 -b(b+1))\,f=0,
\end{equation}
where the variable $u$ is related to $\rho$ by the equation: 
$u=(1-\tanh\rho)/2$.
The above equation is a hypergeometric differential equation and 
its general solution  
is a linear combination of two hypergeometric functions
(cf.~\cite{aands65}, pp. 562 and 563), {\it i.e.} 
\begin{eqnarray}
f(u) &=&  A\; F\left(b+{1\over 2}+c, b+{1 \over 2}-c, 
1+a+b, u\right)\nonumber\\
& &\quad\quad\quad\quad\quad
+\; B\; u^{-a-b}\, F\left({1 \over 2}-a+c, {1 \over 2}-a-c, 1-a-b, u\right),
\end{eqnarray}
where $A$ and $B$ are arbitrary constants and
\begin{equation}
c={\left({1\over 4}+ q^2 {B_0}^2 L^4\right)}^{1/4}.
\end{equation}

To calculate the tunnelling probability for the effective Schr{\" o}dinger
equation~(\ref{eqn:effsch}), we have to choose the constants $A$ and 
$B$ such that $\psi\sim e^{ik_R x}$ as $x \rightarrow +\infty$
({\it i.e.} when $u\rightarrow 0$).
This can be achieved by setting $A=0$ and $B=2^{-b}$, so that
\begin{equation}
f(u)= 2^{-b}\, u^{-a-b}\; F\left({1 \over 2}-a+c, {1 \over 2}-a-c, 
1-a-b, u\right).
\end{equation}
Substituting the above solution in~(\ref{eqn:psi}) and using the 
relation~(cf.~\cite{aands65}, p. 559)
\begin{eqnarray}
& & F\left({1 \over 2}-a + c, {1 \over 2}-a-c, 1-a-b, u\right)\nonumber\\
& &\quad\quad\quad
=\, P\, F\left({1 \over 2}-a+c, 
{1 \over 2}-a-c, 1-a+b, 1-u\right)\nonumber\\
& &\quad\quad\quad\quad\quad
+\, Q\,(1-u)^{a-b}\;
F\left({1 \over 2}-b-c, {1 \over 2}-b+c, 1+a-b, 1-u\right),\quad
\end{eqnarray}
where
\begin{equation}
P = \left({\Gamma(1-a-b)\; \Gamma(a-b) \over 
{\Gamma\left({1 \over 2}-b-c\right)
\Gamma\left({1 \over 2}-b+c\right)}}\right)\;\; {\rm and}\;\;
Q = \left({\Gamma(1-a-b)\; \Gamma(b-a) \over 
{\Gamma\left({1 \over 2}-a+c\right)
\Gamma\left({1 \over 2}-a-c\right)}}\right),
\end{equation}
we find that, as $x \rightarrow -\infty$, {\it i.e} when 
$(1-u)\rightarrow 0$, 
\begin{equation}
\psi\rightarrow P\, e^{ik_L x} + Q\, e^{-ik_L x}.\label{eqn:psiR}
\end{equation}
Consider a solution of the effective Schr{\" o}dinger 
equation~(\ref{eqn:effsch}) which goes as $\left(Re^{ik_L x} 
+e^{-ik_L x}\right)$ as $x\rightarrow -\infty$ and goes over to 
$\left(Te^{ik_R x}\right)$ as $x\rightarrow +\infty$. 
Then it is easy to identify the expressions 
for $R$ and $T$ from equation~(\ref{eqn:psiR}).  
They are given by
\begin{eqnarray}
R &=& \left({P \over Q}\right)
= \left({{\Gamma\left({1 \over 2}-a+c\right)
\Gamma\left({1 \over 2}-a-c\right) 
\Gamma\left(a-b\right)}\over 
{\Gamma\left({1 \over 2}-b-c\right)
\Gamma\left({1 \over 2}-b+c\right)
\Gamma\left(b-a\right)}}\right),\nonumber\\
T &=& \left({1\over Q}\right) 
= \left({{\Gamma\left({1 \over 2}-a+c\right)
\Gamma\left({1 \over 2}-a-c\right)} \over {\Gamma(1-a-b)\; 
\Gamma(b-a)}}\right);
\end{eqnarray}
so that
\begin{equation}
{\vert R \vert}^2 =
\left({\cosh2\pi k_{+}L + \cos 2\pi c
\over {\cosh 2\pi k_{-}L + \cos 2\pi c}}\right)
\end{equation}
and
\begin{equation}
{\vert T \vert}^2 = \left({k_L\over k_R}\right)\,
\left({\cosh 2\pi k_{+}L - \cosh 2\pi k_{-}L
\over {\cosh 2\pi k_{-}L + \cos 2\pi c}}\right).\label{eqn:TT}
\end{equation}
The Wronskian condition for the effective Schr{\" o}dinger 
equation~(\ref{eqn:effsch}) then leads us to the 
following relation
\begin{equation}
{\vert R \vert}^2-\left({k_R \over k_L}\right)\,{\vert T \vert}^2
=1.\label{eqn:kp}
\end{equation}
So, the tunnelling probability {\it is} non-zero for the time independent 
magnetic field we have considered here.
It is, in fact, given by ${\vert T\vert}^2$ in equation~(\ref{eqn:TT}). 

The implications of our analysis are discussed in the following section.
 
\section{Conclusions}\label{sec:concl}

A time independent magnetic field does not give rise to an electric 
field and a pure magnetic field cannot do any work on charged 
particles. 
Therefore it seems plausible that such a magnetic field does not 
produce particles.
This expectation is, in fact, corroborated by the result we have obtained 
in section~\ref{sec:efflag}, {\it viz.} that the imaginary part of the 
effective lagrangian for a time independent, but otherwise arbitrary, 
magnetic field is zero.
Our analysis in sections~\ref{sec:efflag} and~\ref{sec:tunnprob} 
has been carried out assuming that the time independent magnetic 
field is described by a space dependent gauge.
In such a gauge, $\beta$ is trivially zero and if we had considered 
only a non-zero $\beta$ to imply particle production, then this result 
would have proved to be consistent with the result we have obtained 
in section~\ref{sec:efflag}.

But this is not the whole story. 
According to the tunnelling interpretation, in a time independent 
gauge it is the tunnelling probability for the effective Schr{\" o}dinger 
equation that has to be interpreted as particle production.
In section~\ref{sec:tunnprob}, we find that there exists a non-zero 
tunnelling probability for a spatially confined and time independent
magnetic field.
If the tunnelling interpretation is true, this result would then imply 
that such a background can produce particles thereby contradicting the 
result we have obtained in section~\ref{sec:efflag}.

The tunnelling probability can, in fact, prove to be non-zero in a 
more general case and is certainly not an artifact of our specific 
example. 
This can be seen as follows:
Consider an arbitrary electromagnetic field described by the vector 
potential
\begin{equation}
A^{\mu}=(\phi(x), 0, A(x), 0),
\end{equation}
where $\phi(x)$ and $A(x)$ are arbitrary functions of $x$.
If the decomposition of the normal modes is carried out as was done
in~(\ref{eqn:nm}), then the effective Schr{\" o}dinger equation for 
the $x$ coordinate corresponding to the above vector potential turns 
out to be 
\begin{equation}
-{d^2\psi \over dx^2}+\left((k_y-qA)^2-(\omega-q\phi)^2\right)\psi
=(-k_z^2-m^2)\,\psi.\label{eqn:phia}
\end{equation}
If we also assume that $\phi(x)$ and $A(x)$ vanish at the spatial
infinities, then it is clear that the solutions for $\psi$ will reduce 
to plane waves as $\vert x\vert \rightarrow \infty$.
When such solutions are possible, in general, there is bound to exist 
a non-zero tunnelling probability for the effective Schr{\" o}dinger 
equation.
Thus, quite generally, the tunnelling interpretation will force us 
to conclude that the electromagnetic field described by the above 
potential produces particles.
In particular, the tunnelling probability will prove to be be 
non-zero even when $\phi=0$ the case which corresponds to a pure 
time independent magnetic field. 
But for such a case, we have shown in section~\ref{sec:efflag} 
that the effective lagrangian is real and hence there can be no 
particle production.
Thus we again reach a contradiction between the results obtained
from the tunnelling interpretation and those obtained from the 
effective lagrangian.

On the other hand, consider the following situation.
If we choose $A(x)$ to be zero and $\phi(x)$ to be non-zero in the 
above vector potential, then such a vector potential will give rise 
to a time independent electric field.
Such an electric field is always expected to produce particles.
But in the space dependent gauge we have chosen here $\beta$ is trivially 
zero and if we consider only a non-zero $\beta$ to imply particle 
production, then we will be forced to conclude that time independent
electric fields will not produce particles! 
It is to salvage such a situation, that the tunnelling interpretation 
has been repeatedly invoked in literature.
But then, our analysis in the last two sections show that tunnelling 
probability can be non-zero even if effective lagrangian has no 
imaginary part!

\noindent  
There appears to be three possible ways of reacting to this 
contradiction.
We shall examine each of them below:

(i) We may begin by noticing that in quantum field theory, there is always 
a tacit assumption that not only the fields but also the potentials 
should vanish at spatial infinities. 
If we take this requirement seriously, we may disregard the results for 
constant electromagnetic fields (the only case for which explicit results 
are known by more than one method!) as unphysical. 
Then we only need to provide a {\it gauge invariant criterion} for particle 
production in electromagnetic fields described by potentials which 
vanish at infinity.

This turns out to be a difficult task, even conceptually. 
To begin with, we do not know how to generalize Schwinger's analysis and 
compute the effective lagrangian for a spatially varying electromagnetic 
field. 
The only procedure available for us to study the evolution of the 
quantum field in such backgrounds are based on the method of normal 
mode analysis where we go on to obtain the tunnelling probability 
${\vert T\vert}^2$. 
But then, the potential term in the effective Schr{\" o}dinger 
equation is not gauge invariant, as can be easily seen from its 
form in equation~(\ref{eqn:phia}).
So the tunnelling interpretation, even if it is adhered to, has the 
problem that it may not yield results that are gauge invariant. 
In fact, the situation is more serious; the entire tunnelling approach can 
be used only {\it after} a particular gauge has been chosen. 
In some sense, the battle has been lost already.

Operationally also, it is doubtful whether the tunnelling approach will 
yield results that are always consistent with the effective lagrangian.
As the analysis in this paper shows, there is at least one case---that of 
a spatially confined magnetic field---for which one can obtain a formal
expression for effective lagrangian and compare it with the results 
obtained from the normal mode analysis. 
These results are clearly in contradiction with each other.

(ii) One may take the point of view that particle production in an 
electromagnetic field is a gauge dependent phenomenon. 
It appears to be a remedy worse than disease and is possibly not acceptable.
In addition to philosophical objections one can also rule out this possibility 
by the following argument. 
We note that in the laboratory we can produce electromagnetic fields by 
choosing charges and current distributions but we have no operational way 
of implementing a gauge.
So, given a particular electromagnetic field, in some region of the 
laboratory, we will either see particles being produced or not. 
It is hard to see where the gauge can enter this result.

This point has some interesting similarities (and differences) with the 
question of particle definition in a gravitational field. 
If we assume that the choice of gauge in electromagnetic backgrounds  
is similar to the choice of a coordinate system in gravity, then one 
would like to ask whether the concept of particle is dependent on the 
coordinate choice. 
People seem to have no difficulty in accepting a coordinate dependence 
of particles (and particle production) in the case of gravity though 
the same people might not like the particle concept to be gauge dependent 
in the case of electromagnetism!
To some extent, this arises from the idea that a coordinate choice is
implementable by choosing a special class of observers, say, while a 
gauge choice in electromagnetism is not implementable in practice.

(iii) Finally, one may take the point of view that tunnelling 
interpretation is completely invalid and one should rely entirely 
on the effective lagrangian for interpreting the particle production. 
In  this approach one would calculate the effective lagrangian for a 
given electromagnetic field (possibly by numerical techniques, say) 
and will claim that particle production takes place only if the effective 
lagrangian has an imaginary part. 
Further one would confine oneself to those potentials which vanish at 
infinity, thereby ensuring proper asymptotic behavior.

This procedure is clearly gauge invariant in the sense that the effective 
lagrangian is (at least formally) gauge invariant. 
Of course, one needs to provide a procedure for calculating  the effective 
lagrangian without having to choose a particular gauge. 
Given such a procedure, we have an unambiguous, gauge invariant criterion 
for particle production for all potentials which vanish asymptotically.
In fact, the effective lagrangian for a spatially varying electromagnetic
background can be formally expressed in terms of gauge invariant quantities 
that involve the derivatives of the potentials and the fields.

This point could also have an interesting implication for gravitational 
backgrounds.
The analogue of a constant electromagnetic background in gravity corresponds 
to spacetimes whose $R_{\mu\nu\rho\sigma}$'s are constants.
The effective action in gravity can then possibly be expressed in terms of 
coordinate invariant quantities constructed from $R_{\mu\nu\rho\sigma}$'s, 
just as it was possible to express the effective lagrangian for 
a constant electromagnetic background in terms of gauge invariant 
quantities involving $F_{\mu\nu}$'s.

Comparing the three choices listed above, it seems that the third one is 
the most reasonable.
Therefore, we conclude that the results obtained from the effective 
lagrangian can be relied upon whereas the tunnelling approach has to 
be treated with caution. 
It is likely, however, that the tunnelling interpretation will prove to 
be consistent with the effective lagrangian approach if we demand that 
an auxiliary gauge invariant criterion has to be satisfied by the 
electromagnetic background before we can attribute a non-zero tunnelling 
probability to particle production.
But it is not obvious as to how such a condition can be obtained from the 
normal mode analysis. 
The wider implications of this result are under investigation.

\section*{\centerline {Acknowledgments}}

\noindent
LSK is being supported by the senior research fellowship of the
Council of Scientific and Industrial Research, India.
The authors would wish to thank Sukanya Sinha, Abhijit Kshirsagar, 
Anil Gangal and Shiv Sethi for discussions.

\end{document}